\documentclass[final]{elsart5p}

\usepackage{graphicx}
\usepackage{rotating}
\usepackage{amssymb}
\usepackage{mathptmx}
\usepackage[numbers]{natbib}
\usepackage{url, hyperref}

\begin{document}

\begin{frontmatter}

\title{Cryogenic Composite Detectors for the Dark Matter Experiments CRESST and EURECA}

\author{S.~Roth\corauthref{cor1}} \ead{sabine.roth@ph.tum.de}, \author{C.~Ciemniak}, \author{C.~Coppi}, \author{F.~v.~Feilitzsch}, \author{A.~G\"utlein}, \author{C.~Isaila}, \author{J.-C.~Lanfranchi}, \author{S.~Pfister}, \author{W.~Potzel}, \and \author{W.~Westphal}
\corauth[cor1]{Corresponding author. Tel.: +49 89 289 12525; fax.: +49/89/289 126 80.}

\address{Physik-Department E15, Technische Universit\"at M\"unchen,\\ James-Franck-Str., D-85748 Garching, Germany}

\received{6 June 2008}
\revised{17 September 2008}
\accepted{29 September 2008}
\begin{abstract}
Weakly Interacting Massive Particles (WIMPs) are candidates for non-baryonic Dark Matter. WIMPs are supposed to interact with baryonic matter via scattering off nuclei producing a nuclear recoil with energies up to a few 10 keV with a very low interaction rate of $\sim$10$^{-6}$ events per kg of target material and day in the energy region of interest. The Dark Matter experiment CRESST (Cryogenic Rare Event Search with Superconducting Thermometers) and the EURECA project (European Underground Rare Event Calorimeter Array) are aimed at the direct detection of WIMPs with the help of very sensitive modularised cryogenic detectors that basically consist of a transition edge sensor (TES) in combination with a massive absorber crystal.  In the CRESST experiment the search for coherent WIMP-nucleon scattering events is validated by the detection of two processes. In the scintillating absorber single crystal, CaWO$_4$, heat (phonons) and scintillation light are produced and detected with two independent cryogenic detectors: a phonon channel and a separate light channel.\\
The development of such cryogenic detectors and the potential ton-scale production are investigated in this paper. To decouple the TES production from the choice of the target material in order to avoid heating cycles of the absorber crystal and to allow pretesting of the TESs, a composite detector design (CDD) for the detector production has been developed and studied. An existing thermal detector model has been extended to the CDD, in order to investigate, understand, and optimize the performance of composite detectors. This extended model, which has been worked out in detail, can be expected to provide a considerable help when tailoring composite detectors to the requirements of various experiments.
\end{abstract}


\end{frontmatter}

\section{Introduction}

\begin{figure*}[!t]
  \begin{center}
   \includegraphics[width=1\textwidth, keepaspectratio]{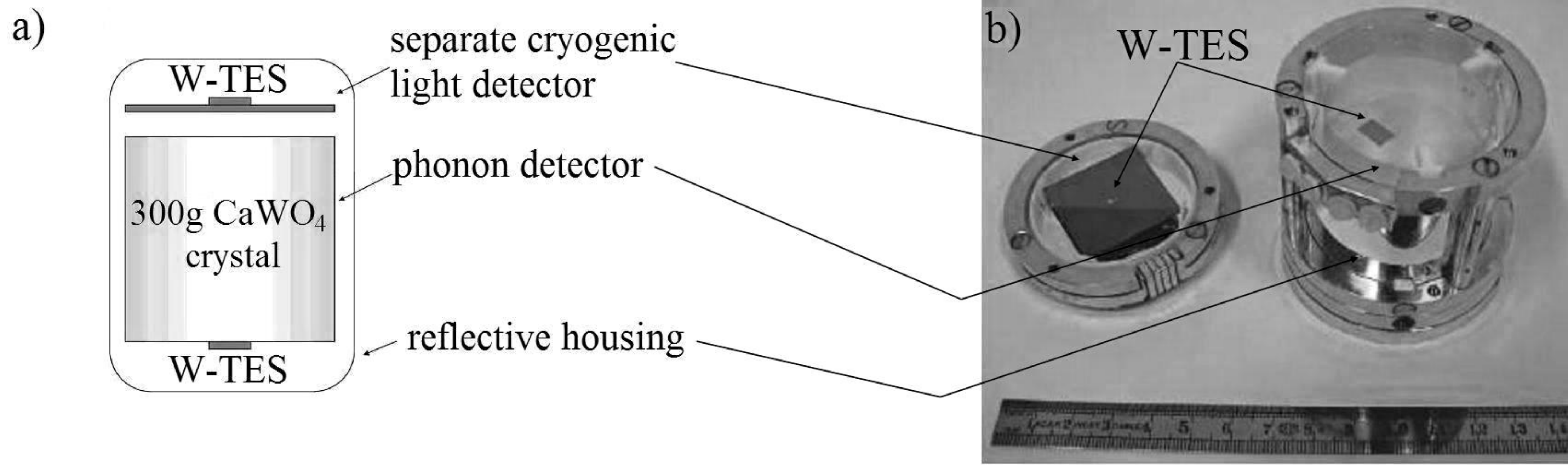}
  \end{center}
  \caption{a) Schematic drawing of a CRESST detector module. b) Photo of a CRESST detector module with (left) the light detector and (right) the phonon detector.}
  \label{CRESSTmodule}
\end{figure*}

The CRESST experiment \cite{CRESSTnew,05CRESST} is aimed at the direct detection of theoretically well motivated Dark Matter (DM) particles - WIMPs \cite{Krauss} (Weakly Interacting Massive Particles) - with the help of sensitive modularised cryogenic detectors which are operated in the low mK-temperature regime. One CRESST detector module consists of a scintillating CaWO$_4$ crystal of $\sim$300 g (height=40 mm, $\oslash$=40 mm) equipped with two independent cryogenic detector channels: the phonon channel, a tungsten transition edge sensor (W-TES) deposited directly onto the surface of the CaWO$_4$ crystal, and the light channel, a SOS (Silicon on Sapphire) or Silicon crystal ($\oslash$=40 mm) also equipped with a W-TES (see Figure \ref{CRESSTmodule}). The detection of WIMPs is expected to occur via coherent WIMP-nucleon scattering in the CaWO$_4$ crystal. In such a process phonons and scintillation light will be produced. The simultaneous measurement of the phonon and the light signal, the so-called phonon-light-technique \cite{Meunier, Jelena Ninkovic}, is used to suppress background caused by electron recoils ($\gamma$ and $\beta$ background), as these produce more light in comparison to the nuclear recoil events by neutrons or WIMPs. However, to efficiently apply this suppression technique and to improve the sensitivity of the experiment to WIMP-recoil events, detectors with a low threshold and good resolution have to be used, while at the same time a big target mass, for future Dark Matter experiments as e.g. EURECA \cite{Eureca} on the tonne scale, is needed.\\
Hence, a highly reproducible and well understood detector design which also allows a large-scale production is required. The chosen design should make it possible to introduce multi-material targets for the WIMP detection. In addition, the decoupling of the W-TES production from the detector crystal is desired, as the scintillation-light yield of the used CaWO$_4$ crystals, as well as of other suitable crystal types, is known to suffer during the deposition process of the TES due to oxygen loss of the crystal \cite{scintillation}. In order to comply with these requirements, the Composite Detector Design (CDD), which was originally developed for the GNO experiment \cite{GNO, DAJL, DATL}, has been further improved (see Section \ref{CDDsec}). For the purpose of optimising and tailoring composite detectors to the needs of DM search experiments, a detailed understanding of the detector response and the signal evolution is required. In this paper we describe a thermal detector model for cryogenic composite detectors \cite{Roth}, which we have developed on the basis of an already existing thermal detector model for cryogenic detectors with the TES directly deposited onto the surface of the absorber crystal \cite{Proebst}.

\section{Cryogenic Composite Detectors}\label{CDDsec}

\subsection{Setup of a Classical Cryogenic Detector}

The cryogenic calorimeters used in the CRESST experiment basically consist of three parts (see Figure \ref{basicdetector1}): 
\begin{figure}[!h]
  \begin{center}
   \includegraphics[width=0.4\textwidth, keepaspectratio]{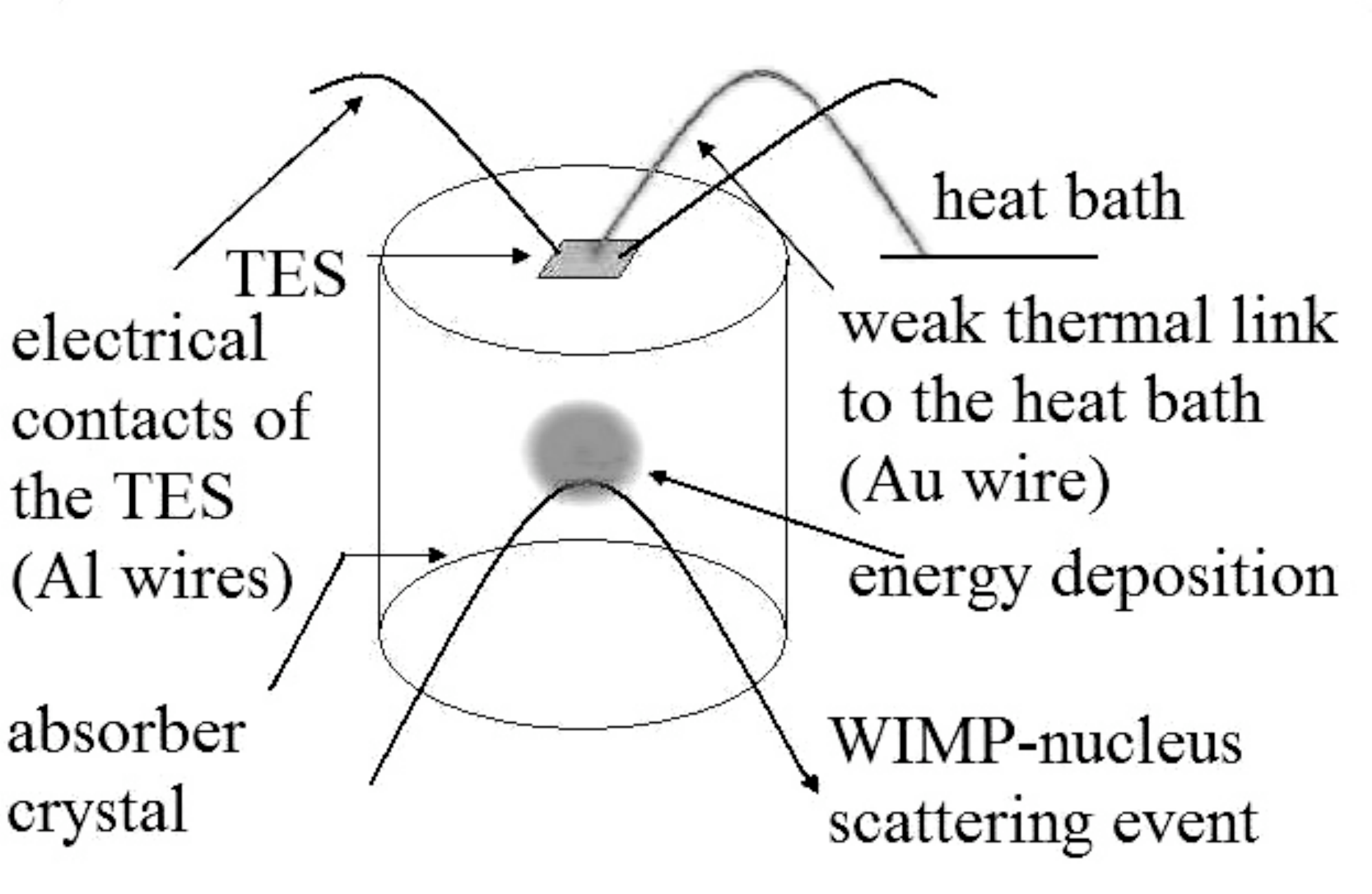}
  \end{center}
  \caption{Sketch of the principle setup of a cryogenic detector.}
  \label{basicdetector1}
\end{figure}
\begin{figure}[!h]
  \begin{center}
   \includegraphics[width=0.4\textwidth, keepaspectratio]{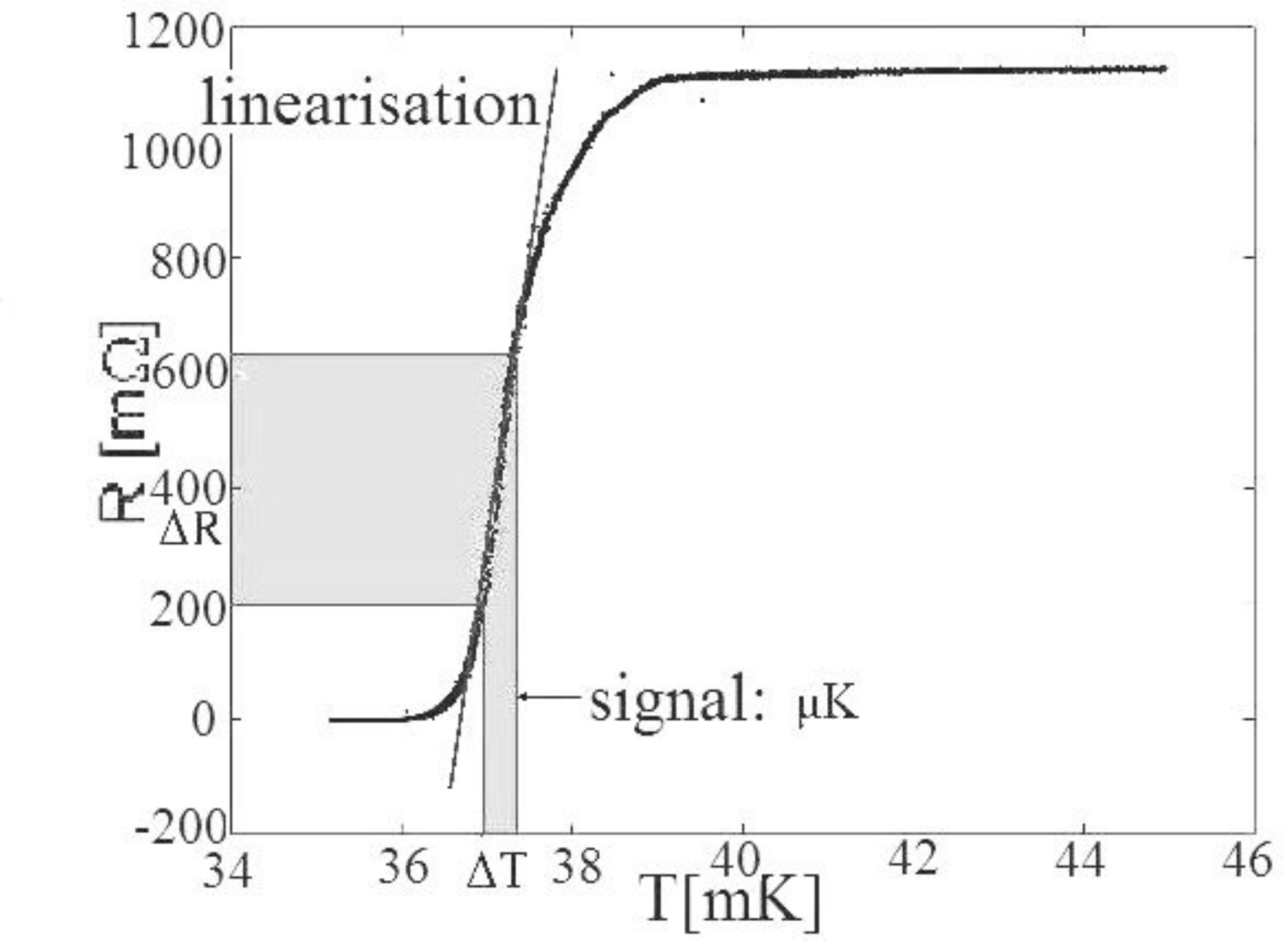}
  \end{center}
  \caption{Working principle of a TES: resistance of a (tungsten) thin film in the transition region from the super to normalconducting state. A small change in temperature $\Delta$T leads to a measurable change in resistance $\Delta$R which is then readout with a DC SQUID circuit \cite{SQUID1, SQUID2}.}
  \label{basicdetector2}
\end{figure}
\begin{figure*}[!t]
  \begin{center}
   \includegraphics[width=1\textwidth, keepaspectratio]{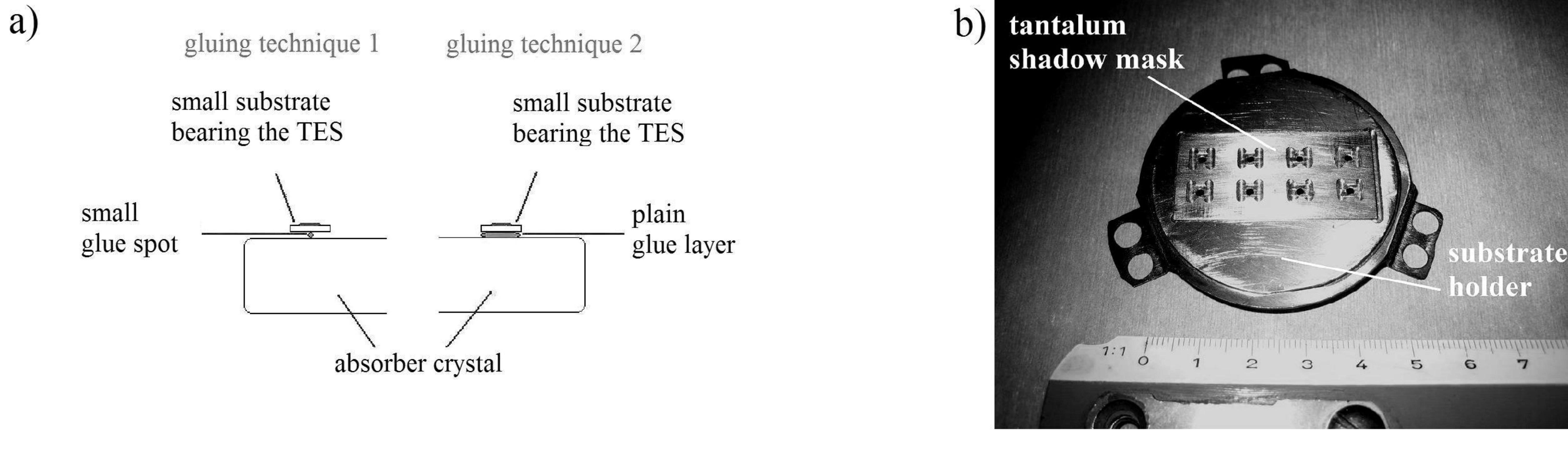}
  \end{center}
  \caption{a) Sketch of the Composite Detector Design with two different gluing techniques, either a small glue spot or a larger glue area is used to couple the small TES-carrier to the larger absorber crystal of the detector. b) Shadow mask used for the deposition of several TESs in one step on small crystal substrates, avoiding the need for photolithographical steps \cite{ChrisitanI}.}
  \label{gluing}
\end{figure*}

\begin{itemize}
\item the absorber crystal (thermally decoupled from its holder), where the energy deposition by a particle causes the production of a phonon population
\item a small metal film, directly deposited onto the absorber crystal surface - the TES \cite{Proebst} (see Figure \ref{basicdetector2}) - which acts as the thermometer
\item a gold wire serving as the thermal link of the system, contacting the detector to the copper holder, in order to allow thermal relaxation of the system after an event
\end{itemize}
During detector operation, the temperature of the TES is stabilized in its transition region (see Figure \ref{basicdetector2}). Hence, a small change in its temperature, as induced by a recoil event in the absorber, leads to a measurable change in its resistance. This resistance change can then be readout with a DC SQUID circuit \cite{SQUID1, SQUID2}.

\subsection{The Composite Detector Design (CDD)}

In contrast to the classical design of a cryogenic calorimeter, for the CDD the TES is produced on a separate small (in comparison to the absorber of the detector) crystal substrate which is afterwards coupled to the absorber by gluing (see Figure \ref{gluing}a). This technique offers several advantages: as the TES-deposition is decoupled from the absorber crystal and material, the absorber crystal is not exposed to any production processes such as heating or etching steps and moreover, the deposition process does not have to be adapted for different materials but can be optimized for only one substrate material. Additionally, using small crystal substrates allows the production of several TESs \cite{ChrisitanI} in one step (see Figure \ref{gluing}b) which enlarges the probability that these TESs exhibit similar properties. Furthermore, given that the TESs are produced separately from the sensitive and expensive absorber crystals, these TESs can be pretested and selected before they are used for manufacturing detectors.\\ 
As can be seen in Figure \ref{gluing}a, two different gluing techniques were introduced: either a very small glue spot or a larger glue area of approximately the size of the substrate (e.g. Al$_2$O$_3$ crystal) or the TES-area was used to attach the TES-carrier to the absorber crystal. The motivations for these two different designs as well as advantages and disadvantages will be discussed in Sections \ref{TDMCCD} and \ref{Conclusions}.

\subsection{Composite Detector with Sapphire Absorber}

\begin{figure*}[!t]
  \begin{center}
   \includegraphics[width=1\textwidth, keepaspectratio]{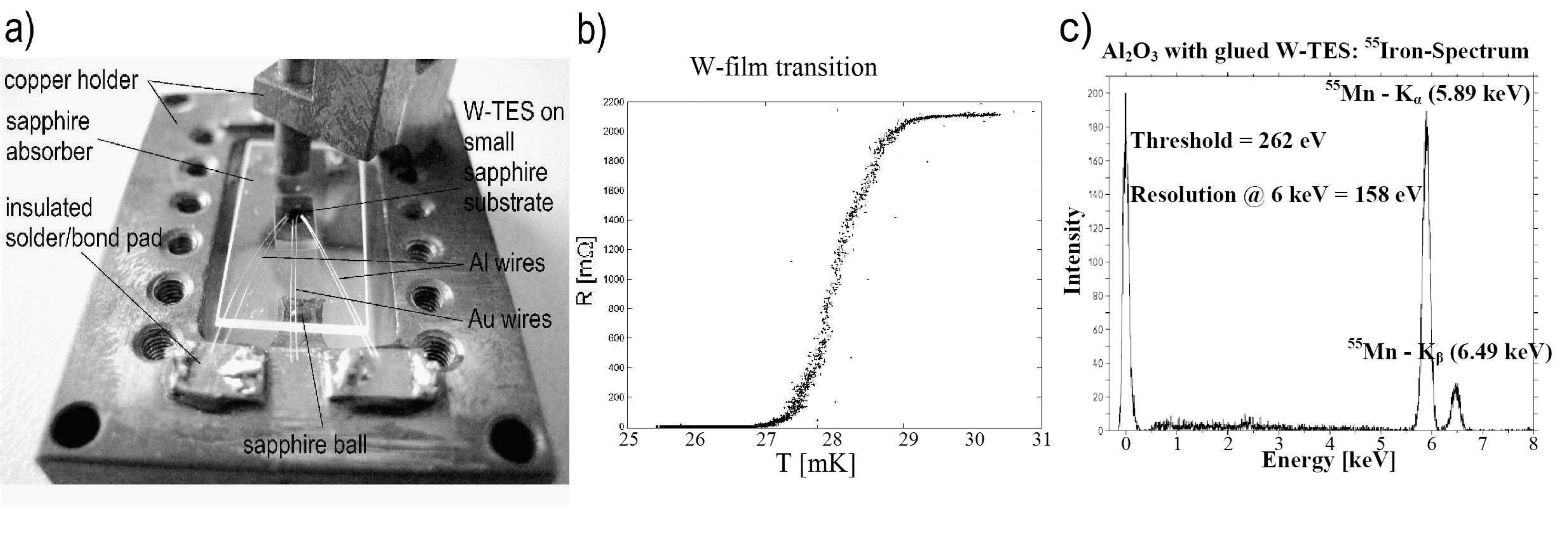}
  \end{center}
  \caption{a) Cryogenic Composite Detector with sapphire absorber. b) Transition of the utilized W-TES. c) $^{55}$Fe-spectrum recorded with the sapphire detector. Separation of the Mn-K$_{\alpha}$ and Mn-K$_{\beta}$ line can clearly be seen.}
  \label{compositedetector}
\end{figure*}

In Figure \ref{compositedetector}a, an example of a realized composite detector using a sapphire absorber of 10$\times$20$\times$1 mm$^3$ is given. The superconducting transition of the utilized W-TES ($\oslash=$2 mm), which was deposited on a sapphire carrier substrate of 5$\times$3$\times$1 mm$^3$, can be seen in Figure \ref{compositedetector}b. A pulse height spectrum recorded with this detector, while it was irradiated with an $^{55}$Fe source, is shown in Figure \ref{compositedetector}c. A quite satisfying resolution of 158 eV @ 6 keV (FWHM) and a threshold of 262 eV could be deduced\footnote{The resolution is derived by taking the full width at half maximum of a gaussian fit to the 5.89 keV peak of the spectrum, while the threshold is given by the 5$\sigma$ width of the peak around 0 keV, which is composed of recorded noise samples of the baseline.}.

\section{Basic Thermal Detector Model}\label{BTDM}

\begin{figure*}[!t]
  \begin{center}
   \includegraphics[width=1\textwidth, keepaspectratio]{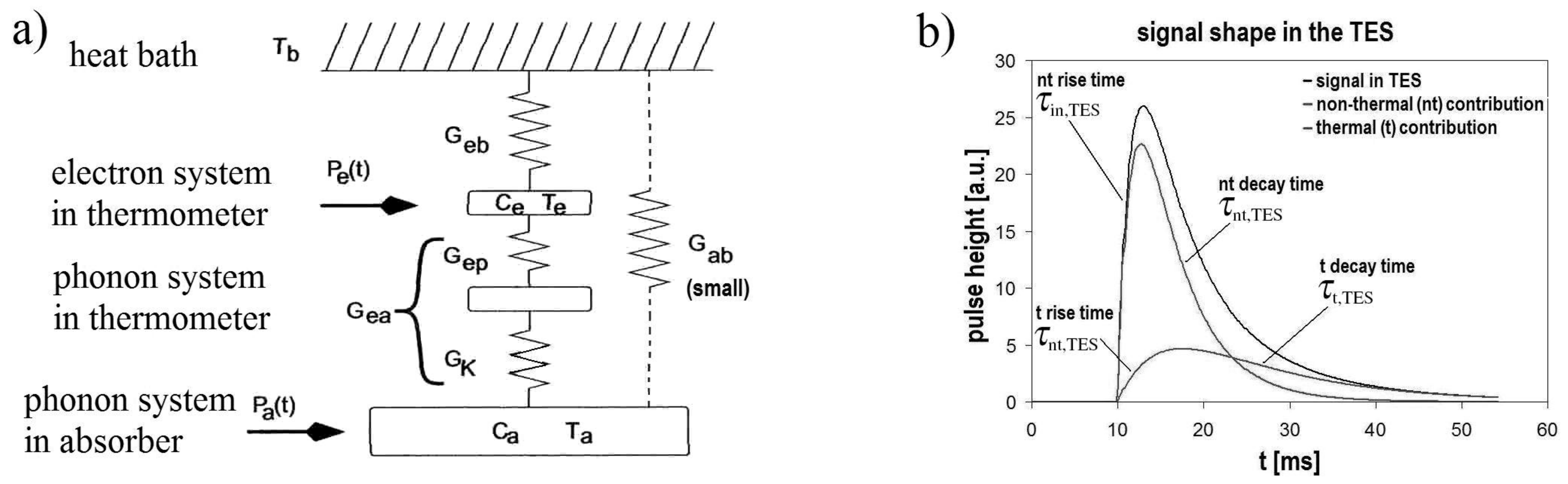}
  \end{center}
  \caption{a) Thermal model of a cryogenic detector with the TES (thermometer) directly deposited on the surface of the absorber \cite{Proebst}. T$_b$ is the heat-bath temperature, T$_e$ and T$_a$ are the temperatures of the TES electron system and the absorber phonon system, respectively, C$_e$ and C$_a$ are their heat capacities. Phonons that thermalise in the electron system of the TES or in the absorber cause the energy input P$_e$ or P$_a$ into the TES and the absorber, respectively. G$_{ab}$, G$_{eb}$ and G$_{ea}$ are thermal conductances.}
  \label{thermaldetectormodel}
\end{figure*}

The response of a classical cryogenic detector to an event, that is, the resulting pulse shape of the signal after an energy deposition, can be depicted with the help of a thermal detector model \cite{Proebst} which describes the system in terms of its thermal components, their heat capacitites C$_i$, and the thermal conductances G$_i$ connecting these components (see Figure \ref{thermaldetectormodel}a).\\
Immediately after an event, the energy deposition in the absorber crystal gives rise to the formation of nonthermal ($nt$) phonons in the absorber. After a very short time (of the order of $\sim$ 10 $\mu$s )\footnote{This nonthermal phonon population represents a higher temperature than the one the system exhibits (thermal equilibrium state) and is spread over the crystal on a time scale much shorter than all other time scales involved in the signal evolution process for the detectors regarded here.}, these $nt$ phonons are uniformly distributed over the whole crystal via a few surface reflections \cite{Proebst}. The properties of the detector and the signal evolution are fundamentally affected by the different relaxation processes of this $nt$ phonon population that are possible in the system. One of the two most important possibilities for the detectors regarded here is the transmission of $nt$ phonons through the absorber-TES interface into the TES itself. Here, with a certain probability, they rapidly thermalise via efficient absorption by the free electrons of the metal. The other possibility taken into account here is the (more slowly) thermalisation of $nt$ phonons in the absorber by inelastic scattering processes at the crystal surfaces. The second process leads to a temperature rise of the absorber and gives rise to a thermal ($t$) phonon population in the absorber.

The absorption process of the $nt$ phonons in the TES-film provides an energy input P$_{\mbox{\tiny{e}}}$(t) into the electronic system of the TES. It takes place on a typical time scale of the $nt$ phonon-collection time $\tau_{\mbox{\tiny{film}}}\propto A_{\mbox{\tiny{TES}}}^{-1}$ (with $A_{\mbox{\tiny{TES}}}$: TES-area) in the TES. The excited electron system in the TES then relaxes quickly via the thermal conductance G$_{eb}$ into the heat bath\footnote{For the detectors regarded here, this thermal conductance G$_{eb}$ is chosen to be large. As a consequence, this TES-relaxation process, characterised by $\tau_{\mbox{\tiny{in,TES}}}$, is the fastest relaxation process in the system, much faster than $\tau_{\mbox{\tiny{t,TES}}}$ and $\tau_{\mbox{\tiny{nt,TES}}}$.} inducing the rise time $\tau_{\mbox{\tiny{in,TES}}}$ of the signal in the TES. This absorption process (thermalisation in the TES) is competing with the thermalisation process in the absorber crystal with a typical time scale of $\tau_{\mbox{\tiny{crystal}}}$, leading to an effective non-thermal phonon lifetime of $\tau_{\mbox{\tiny{nt,TES}}}=(\tau_{\mbox{\tiny{film}}}^{-1}+\tau_{\mbox{\tiny{crystal}}}^{-1})^{-1}$ in the system. Thus, the $nt$ decay time $\tau_{\mbox{\tiny{nt,TES}}}$, representing the decay of $nt$ phonons into $t$ phonons, essentially determines the rise time of the $t$ contribution (see Figure \ref{thermaldetectormodel}b). Hence, the absorption process in the TES provides a contribution to the signal in the TES (nt contribution) with a decay time of $\tau_{\mbox{\tiny{nt,TES}}}$. The $t$ phonon population created in the absorber leads to a temperature rise of the TES. By transmission of the heating power via the thermal conductances G$_i$, on a typical time scale $\tau_{\mbox{\tiny{t,TES}}}$ (= lifetime of $t$ phonons in the absorber), a rather slow decay of the $t$ signal contribution is generated. These processes result in a signal shape in the TES as can be seen in Figure \ref{thermaldetectormodel}b, given by equation (\ref{signalshape}) with $A_{nt}$ and $A_{t}$ representing the amplitudes of the two partial pulses and $\Theta(t)$ being the step function, defining the starting point of the pulse:
\begin{figure*}[!t]
  \begin{center}
   \includegraphics[width=1\textwidth, keepaspectratio]{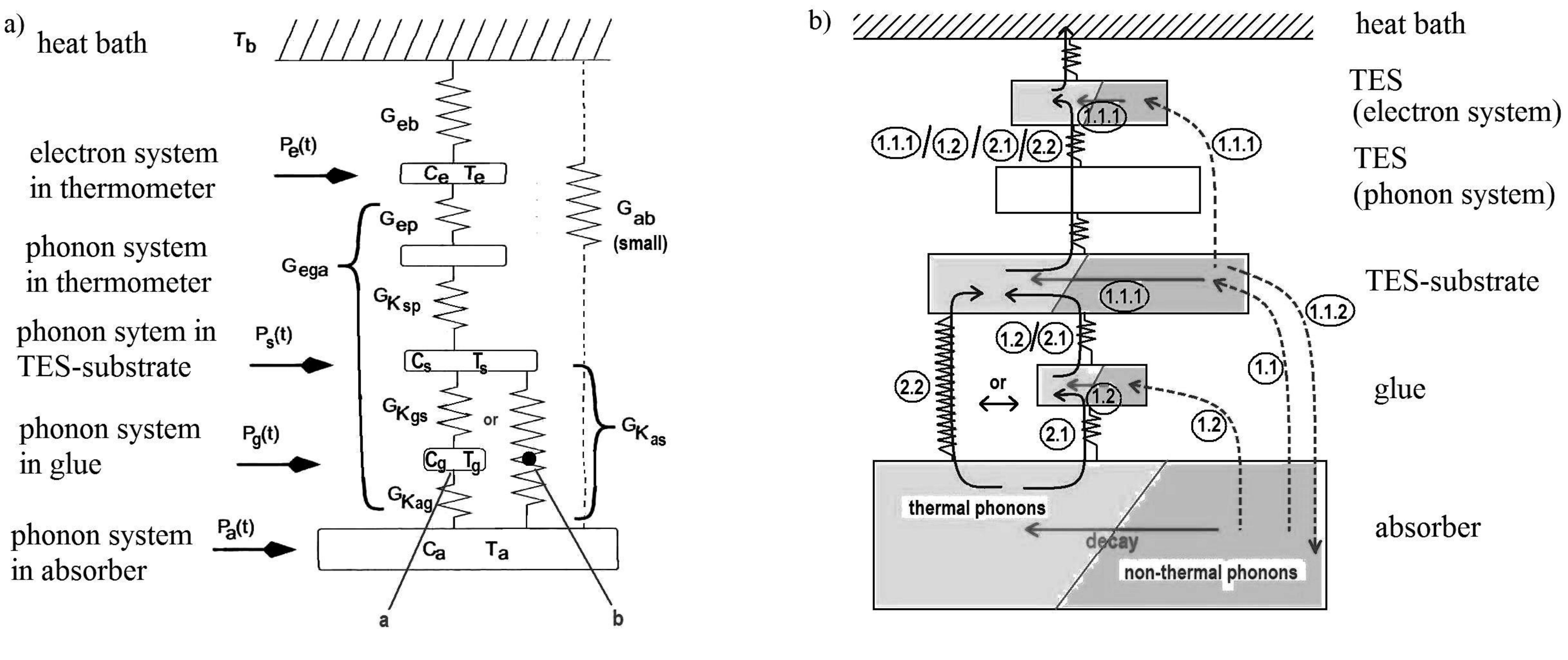}
  \end{center}
  \caption{a) Thermal model of a Cryogenic Composite Detector in terms of its thermal components with two different possibilities of including the glue (\textit{a} as thermal component or \textit{b} as contribution to the thermal conductance). b) Sketch of the most important possibilities of evolving thermal ($t$) and nonthermal ($nt$) phonon populations and their further propagations in the system. Dashed arrows indicate propagation of $nt$ phonons; full arrows indicate propagation of $t$ phonons. The horizontal arrows within the components of the system depict the decay of $nt$ phonons into $t$ phonons. The contribution of the glue is indicated in two ways: it is either considered as a thermal component with a heat capacity and a temperature, where $nt$ phonons can enter and decay or it contributes just a thermal conductance to the system. The numbers given correspond to the different phonon propagations discussed in detail in Ref.  \cite{Roth}.}
  \label{gluetdmphonons}
\end{figure*}

\begin{eqnarray}
& \Delta T(t)= \quad \quad \quad \quad \quad \quad \quad \quad \quad \quad \quad \quad \quad \quad \quad \quad \quad \quad \quad\label{signalshape}\\
& \Theta(t)\left[A_{nt} (e^{-\frac{t}{\tau_{\mbox{\tiny{nt,TES}}}}}-e^{-\frac{t}{\tau_{\mbox{\tiny{in,TES}}}}}) + A_{t} (e^{-\frac{t}{\tau_{\mbox{\tiny{t,TES}}}}}-e^{-\frac{t}{\tau_{\mbox{\tiny{nt,TES}}}}})\right] \nonumber
\end{eqnarray}

\section{Thermal Detector Model for Cryogenic Composite Detectors}\label{TDMCCD}

\subsection{Graphical Representation of a Cryogenic Composite Detector in the Thermal Detector Model}
In order to describe a Composite Detector with the help of a Thermal Detector Model, the glue and the TES-substrate have to be included into the model and their influence on the evolution and propagations of the phonon populations has to be investigated. In Figure \ref{gluetdmphonons}a, a graphical representation describing a cryogenic composite detector in terms of its thermal components and conductances is given.\\
In this picture two principle possibilities of including the glue into the model are indicated: path \textit{a}, the glue is included in the system as thermal component with a heat capacity, where $nt$ phonons can enter and decay, inducing a temperature rise, i.e. , generating a $t$ phonon population in the glue; path \textit{b}, the glue provides just a thermal conductance between the absorber crystal and the TES-substrate.

\subsection{Thermal Detector Model for Cryogenic Composite Detectors}
\begin{figure*}[!t]
  \begin{center}
   \includegraphics[width=1\textwidth, keepaspectratio]{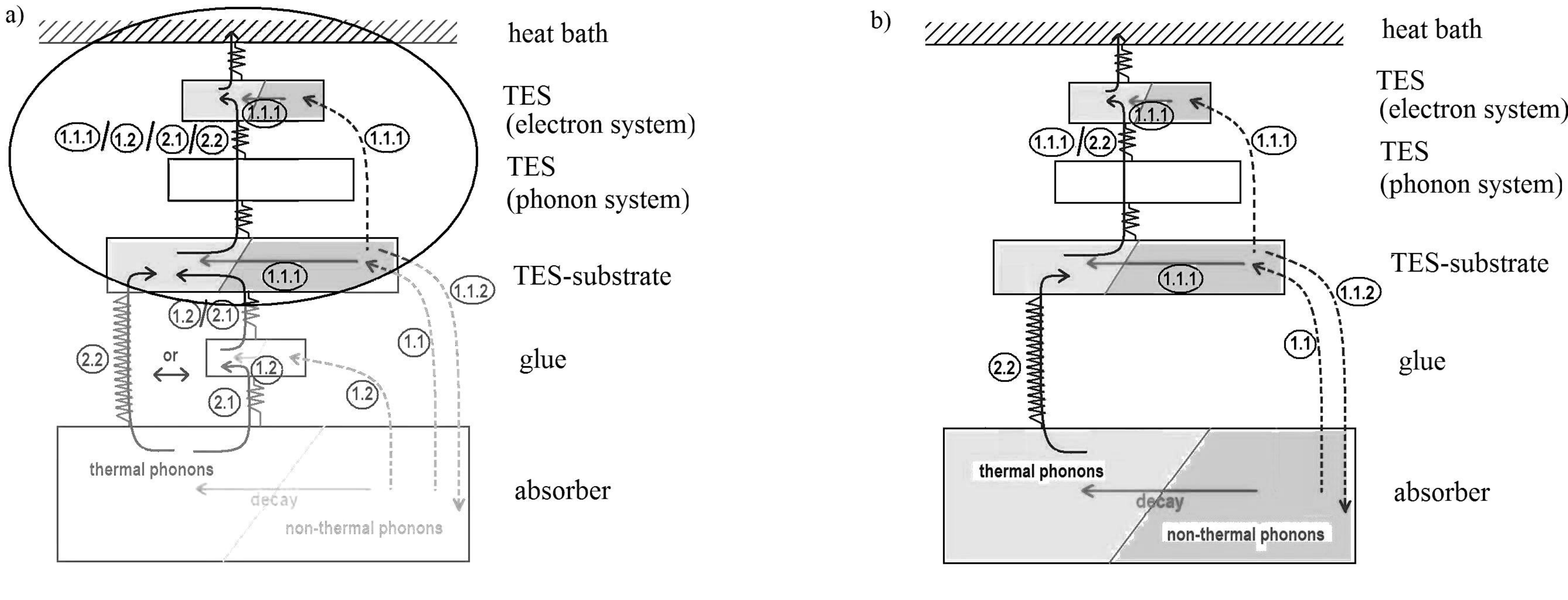}
  \end{center}
  \caption{a) The \textit{TES-detector system} can be treated as a classical cryogenic detector with a slow energy input. b) Dominant $nt$ and $t$ phonon contributions to the signal, experimentally identified by the pulse shape analyses of three different composite detectors.}
  \label{TES-detector}
\end{figure*}

In such a system many different possibilities for the evolution of phonon populations and their propagations emerge (see Figure \ref{gluetdmphonons}b). After an event in the absorber, a $nt$ phonon population builds up there (compare Section \ref{BTDM}) and can again decay in the absorber crystal, giving rise to a $t$ phonon population there. These $t$ and $nt$ phonon populations from the absorber crystal can then propagate through the glue and the TES-substrate into the TES, where in each of these components the $nt$ phonons can create (via their decay) additional $t$ phonon populations.\\
In order to identify the dominant contributions to the signal in a Composite Detector, the possible contributions of the different phonon populations have to be calculated and then compared with the results of the pulse shape analysis of composite detectors with various well defined absorber, TES and glue areas \cite{Roth}.\\
The basic concept which is used for the analysis is demonstrated in Figure \ref{TES-detector}a. Focusing on the upper part of a composite detector, the \textit{TES-substrate - TES system}, it can be recognized that this system can be treated as an individual classical cryogenic detector which can be described by the Basic Thermal Detector Model (see Section \ref{BTDM}). However, the energy input into this \textit{TES-detector system} is no longer delivered by an event in its "absorber", the \textit{TES-substrate}, but is introduced by a slow ($\gg$ 10 $\mu$s, in comparison to the direct event in an absorber of a classical detector) energy input of $t$ and $nt$ phonons ($P_{\mbox{\tiny{nt}}}$ and $P_{\mbox{\tiny{t}}}$) from the absorber crystal via the glue area.\\

Thus, to calculate the possible signal contributions and their respective rise and decay times, the power inputs from the absorber ($P_{\mbox{\tiny{nt}}}$ and $P_{\mbox{\tiny{t}}}$) have to be convoluted with the \textit{TES-detector} signal function $\Delta T(t)$ (see Eq.(\ref{signalshape})). The shape of the power inputs, their rise and decay times are given by the generation times and lifetimes of the respective phonon populations: the $nt$ phonon generation time is again very short ($\sim$ 10 $\mu$s) and can be neglected. The $nt$ phonon lifetime is the combination of the thermalisation time in the absorber crystal and the collection time in the glue area. The $t$ phonon generation time is determined by the lifetime of the $nt$ phonons while the lifetime of the $t$ phonons is defined by the heat capacities of the components and the size of the thermal conductances between them\footnote{For example, to calculate the resulting signal shape for the path \textit{1.2} (see Figure \ref{gluetdmphonons}b and Figure \ref{TES-detector}), first the lifetime of $nt$ phonons in the glue, that is, the thermalisation time in the glue, has to be convoluted with the $nt$ phonon flow $P_{\mbox{\tiny{nt}}}$ into the glue. This results in the generation time of the produced $t$ phonon population in the glue and therefore in the rise time of the resulting $t$ power input into the TES-substrate.}. For the case that the glue is transparent for $nt$ phonons (path \textit{1.1} in Figure \ref{gluetdmphonons}b), three different further propagation possibilities for $nt$ phonons that have reached the TES-substrate have to be taken into account.  As can be expected in analogy with the Basic Thermal Detector Model (see Section \ref{BTDM}), these $nt$ phonons can either be absorbed in the TES (dashed arrow \textit{1.1.1}) or decay into $t$ phonons by thermalisation in the TES-substrate (horizontal arrow \textit{1.1.1}). However, now also the competing process of the propagation \textit{back} into the absorber crystal (dashed arrow \textit{1.1.2}) has to be taken into account.

\subsection{Results}\label{results}

The dominant signal contributions in a Composite Detector and the underlying mechanisms could successfully be identified by comparing the results of these calculations to the pulse shapes of three realized composite detectors, which were built using similar TES-substrates and either different absorbers and the same gluing areas or vice versa, similar absorbers but different gluing areas. As can be seen from Figure \ref{TES-detector}b, an important conclusion of the model is that the glue only has to be included as a thermal conductance between the absorber and the TES-substrate, but not as a thermal component of the system\footnote{That is, for the detectors regarded here, where the volume of the glue is much smaller than the volumes of the absorber and the substrate.}. This can be derived from the result that no dominant contribution from a thermalisation of $nt$ phonons in the glue can be observed. Figure \ref{TES-detector}b also shows that only two dominant possibilities (path \textit{1.1}: $nt$ contribution and \textit{2.2}: $t$ contribution) remain for phonons from the absorber to create a signal in the TES. This implies that the glue is, in principle, transparent for $nt$ and $t$ phonons. However, depending on the exact detector design (i.e. the \textit{TES-to-glue-area} ratio) either only the $nt$ phonons (along path \textit{1.1}) or both phonon types contribute significantly to the signal in the TES. Hence, the model predicts, that in every composite detector a $nt$ phonon contribution from the absorber can be detected, delivering one decay time $\tau_{\mbox{\tiny{nt,as}}}$ (from the \textit{a}bsorber into the TES-\textit{s}ubstrate) only.\\
This implies that $nt$ phonons enter the TES-substrate through the glue area. Different dominant possibilities for their further propagation and evolution are present: the decay within the TES-substrate \cite{Roth}, the absorption in the TES via the \textit{TES-area} (resulting in a signal contribution with decay time $\tau_{\mbox{\tiny{nt,as}}}$) and, controlled by the \textit{TES-to-glue-area} ratio, also the propagation back into the absorber via the \textit{glue-area}. The propagation back into the absorber is, however, only a dominant possibility, when the \textit{TES-to-glue-area} ratio is smaller than 1, i.e., if the glue-area delivers a competitive escape possibility for $nt$ phonons in comparison to the TES-area. In this case, $nt$ phonons, that entered the TES-substrate are not efficiently removed from the detector system by absorption in the TES, but are partially scattered back into the absorber, where they can generate a significant $t$ phonon population, which, in turn, is then also dominantly contributing to the signal in the TES, delivering a second decay time $\tau_{\mbox{\tiny{t,as}}}$.\\
In summary, it can be stated, that a \textit{TES-to-glue-area} ratio larger than 1 leads to only one fast $nt$ significant signal contribution in the TES (one decay time, $\tau_{\mbox{\tiny{nt,as}}}$), while a \textit{TES-to-glue-area} ratio smaller than 1 induces two different significant signal contributions, a fast $nt$ and a slower $t$ phonon contribution with the decay times, $\tau_{\mbox{\tiny{nt,as}}}$ and $\tau_{\mbox{\tiny{t,as}}}$. (For further details see \cite{Roth}.)

\section{Discussion}\label{Conclusions}

With the Thermal Detector Model for Cryogenic Composite Detectors that is presented in the previous section, the signal shape of the three composite detectors investigated can be explained consistently. It has been found that a glue area smaller than the TES-area (see gluing technique 1 in Figure \ref{gluing}a) just offers a small collection area for the $nt$ phonons\footnote{The $nt$ phonon contribution is always the faster one for the detectors investigated here.} into the TES-substrate, while on the other hand, it enlarges the ratio of $nt$ to $t$ phonon contributions to the signal, as discussed in Section \ref{results}. A ratio of $nt$ to $t$ phonon contributions to the signal as large as possible, is strived for in order to improve the resolution and the threshold of the detector. This can be explained by the fact that the larger and the faster this $nt$ contribution, the more efficiently the deposited energy is collected in the TES and therefore creates a more distinct and higher signal in the TES (see Figure \ref{thermaldetectormodel}b). On the one hand, a faster and more efficient collection of the $nt$ phonons into the TES can be achieved with a larger glue area, delivering a larger transmission probability for the $nt$ phonons from the absorber into the TES-substrate. On the other hand, using a larger glue area, involves the introduction of a larger mass of glue, that is, a larger heat capacitiy into the system and additionally an increased risk of introducing spatial inhomogeneities into the system, which in turn could worsen the resolution of the detector. Furthermore, if the glue-area is chosen larger than the TES-area, part of the energy deposited is detected in terms of the slower $t$ signal contribution, as discussed in Section \ref{results}, which prolongs the signal, that is, smears out the signal height and thus worsens the energy information contained in the pulse height.\\
Hence, for the purpose of enlarging the $nt$ phonon contribution to the signal, a rather large, desirably perfectly homogeneous glue area (see gluing technique 2 in Figure \ref{gluing}a) should be used in combination with a \textit{TES-to-glue-area} ratio larger than 1. For this reason, the TES-area should be larger than the glue-area. However, the TES-area cannot be enlarged arbitrarily, for the following reason: A larger TES-area, in turn, introduces a larger heat capacity of the TES. This larger heat capacity would induce a smaller height of the signal for the same amount of energy deposited in the TES. Thus, a trade-off between a large glue area (= $nt$ phonon collection area, enlarged pulse height of the system) and a TES area, most desirably larger than this glue area (= enlarged $nt$ relative to $t$ phonon signal contribution, however, large heat capacitiy), has to be aimed at.\\
One possibility, that could be realized for Dark Matter detectors, e.g. for the CRESST and EURECA experiments, to further improve the $nt$ phonon collection efficiency would be the introduction of superconducting phonon collectors \cite{Stark, WWPHd} on the TES-substrate. Using this technique, a \textit{TES-to-glue-area} ratio larger than 1 and a large $nt$ phonon collection area could be achieved at the same time, without the need for enlarging the TES itself and therefore, without introducing a large additional heat capacity.

\section{Conclusions}


The CRESST Dark Matter experiment  uses scintillating single crystals as targets of cryogenic detectors for Weakly Interacting Massive Particles (WIMPs). The Composite Detector Design (CDD), presented in this paper, can be applied to build cryogenic phonon and light detectors. The CDD enables their mass production as well as a careful treatment of the sensitive absorber crystals with respect to the scintillation light yield. Within the CDD, the transition edge sensor (TES) production is decoupled from the optimization of the scintillation properties of the absorber crystal. This is realized by the application of a gluing technique to attach the TES to the absorber. A Thermal Detector Model for Cryogenic Composite Detectors can consistently explain the influence of this gluing technique on the signal evolution and signal shape. Depending on the \textit{TES-to-glue-area} ratio,  the signal is composed either of one dominant fast nonthermal phonon contribution, or of a fast nonthermal and an additional (slower) thermal contribution. For a good detector performance with respect to energy resolution and threshold, a nonthermal phonon contribution as large as possible is strived for. The introduction of phonon collectors on the TES-substrate is suggested to keep the TES-area, i.e. the introduced heat capacity, as small as possible. To optimize the detector design and to decide on the exact detector parameters, the developed thermal model, and the resulting improved understanding of the signal evolution in a Composite Detector, should be of considerable help.

\ack
This work has been supported by funds of the DFG (SFB 375, Transregio 27: "Neutrinos and Beyond"), the Munich Cluster of Excellence ("Origin and Structure of the Universe"), the EU network for Applied Cryogenic Detectors (HPRN-CT2002-00322), and the \-Maier-Leibnitz-Laboratorium (Garching).

\end{document}